\documentclass[pra, twocolumn, showpacs, nofootinbib, preprintnumbers]{revtex4}
\usepackage{amsmath}
\usepackage{graphicx}

\newcommand{\del}[2]{\ensuremath{\frac{\partial^{#1}}%
{\partial#2^{#1}}}}
\newcommand{\inv}{^{-1}}

\newcommand{\<}{\left<} 
\renewcommand{\>}{\right>}

\begin{document}

\title{Hydrogen-atom spectrum under a minimal-length hypothesis}

\author{S\'andor Benczik} 
\author{Lay Nam  Chang} 
\author{Djordje Minic} 
\author{Tatsu Takeuchi}

\affiliation{Institute
  for Particle Physics and Astrophysics, Virginia Tech, Blacksburg,
  Virginia 24061, USA}

\date{\today}

\begin{abstract}
  The energy spectrum of the Coulomb potential with minimal length
  commutation relations $[X_i, P_j] = i\hbar\{\delta_{ij}(1+\beta P^2)
  + \beta'P_iP_j\}$ is determined both numerically and perturbatively
  for arbitrary values of $\beta'/\beta$ and angular momenta $\ell$.
  The constraint on the minimal length scale from precision hydrogen
  spectroscopy data is of order of a few GeV$\null^{-1}$, weaker than 
  previously claimed.
\end{abstract}

\preprint{VPI-IPPAP-05-01}

\maketitle

Quantum gravity incorporates Newton's constant as a dimensional
parameter that could manifest itself as a minimal length in the
system.  Recent string theoretic considerations suggest
that this length scale might imply an ultraviolet-infrared (UV-IR)
correspondence, contrary to the normal perceptions on momentum and
spatial separations.  Large momenta are now directly tied to large
spatial dimensions, which then implies the existence of a minimal
length.  Earlier studies have focused upon its amelioration of
ultraviolet divergences \cite{earlyQM}, but did not take into full
account the UV-IR correspondence.

There are various ways of implementing such an idea, but the simplest is to
suppose that coordinates 
no longer commute in  $D$-dimensional space.  This, in turn,
leads to a deformation of the canonical commutation relations.  In our
previous works, we adopted the equivalent hypothesis that the fundamental
commutation relations between position and momentum are no longer
constant multiples of the identity.  In this paper, we report on
constraints on the minimal length hypothesis from precision
measurements on hydrogenic atoms.   This system has a potential that is
singular at the origin, and is therefore particularly sensitive to whether
there is a fundamental minimal length.  Considerations based upon 
higher-dimensional theories suggest that such lengths may be large \cite{LXD}.

To set the context, we note that if in one dimension we have
\begin{equation}
  \label{MLCR1D}
  [\hat X, \hat P] = i\hbar(1+\beta \hat P^2) ,
\end{equation}
where $\beta$ is a small parameter, then the resulting uncertainty relation
$
  \label{MLUR}
  (\Delta X)(\Delta P) \ge i\hbar\{1+\beta (\Delta P)^2\}
$
exhibits a form of the UV-IR correspondence, and gives as minimal length
$
  \Delta X \ge \hbar\sqrt{\beta}
$
\cite{Kempf}. 

We had examined the harmonic oscillator system under this hypothesis in
\cite{HO}, but no real constraint can be obtained on
the minimal length, presumably because of the softness of the potential 
at the origin. An interesting approach is to take the classical 
limit $\hbar \to 0$ of the commutation relations; it
yields an unbelievably strong bound, but its robustness might be
questioned \cite{ClassLim}.

We will work in arbitrary $D>1$ dimensions, where \eqref{MLCR1D} takes
the tensorial form
\begin{equation}
  \label{MLCR}
  [\hat X_i, \hat P_j] = i\hbar\{\delta_{ij}(1+\beta \hat P^2) + 
  \beta' \hat P_i \hat P_j\} ,
\end{equation}
which, assuming that the momenta commute
$
  [\hat P_i, \hat P_j] = 0 ,
$
leads via the Jacobi identity to the nontrivial position
commutation relations
\[
  [\hat X_i, \hat X_j] = i\hbar\,
  \frac{(2\beta-\beta')+ (2\beta+\beta')\beta \hat P^2 }
  { (1+\beta \hat P^2) }
  ( \hat P_i \hat X_j - \hat P_j \hat X_i ).
\]

The position and momentum operators can be represented by
\begin{equation}
  \label{repr}
  \hat X_i = (1+\beta \hat p^2) \hat x_i
  + \beta' \hat p_i \hat p_j \hat x_j, \qquad
  \hat P_i = \hat p_i,
\end{equation}
where the operators $\hat x_i$ and $\hat p_j$ satisfy the canonical
commutation relations
$
  [\hat x_i, \hat p_j] = i\hbar\delta_{ij}.
$
The simplest representation is momentum diagonal, 
\begin{equation}
  \hat x_i = i\hbar\del{}{p_i}, \qquad
  \hat p_i = p_i .
\end{equation}


In this representation the eigenvalue equation for the distance squared 
operator $\hat R^2=\hat X_i \hat X_i$ can be solved exactly. With
\begin{equation}
  \label{fromHere}
  z= \frac{(\beta+\beta')p^2 -1}{(\beta+\beta')p^2 + 1},
\end{equation}
the eigenvalues $ {r_{n\ell}^2} = \hbar^2(\beta+\beta')\rho_{n\ell}^2$ 
and eigenfunctions $R_{n\ell}$ are given by  (see \cite{HO} for details)
\begin{gather}
  \rho_{n\ell}^2 = 
  (2n+a+b+1)^2
  - (1-\eta)^2
  \left( L^2 + \dfrac{ (D-1)^2 }{ 4 } \right) ,
  \notag\\
R_{n\ell}(z) 
  \propto
  (\beta+\beta')^{D/4} 
  \left.\left(\frac{1-z}2\right)\!\!\right.^{\lambda/2}
  \left.\left(\frac{1+z}2\right)\!\!\right.^{\ell/2}
  P_n^{(a,b)}(z) ,
  \notag
\end{gather}
where $P_n^{(a,b)}(z)$ are the Jacobi polynomials and
\begin{align}
  \eta&=\frac{\beta}{\beta+\beta'}, \qquad
  &a&=\sqrt{ \frac{ [1 + (D-1)\eta]^2 }{ 4 } + \eta^2 L^2 }, \notag\\
  b&=\frac{D}{2} + \ell - 1, \qquad
  &\lambda &=\frac{1 + (D-1)\eta}{2} + a .
\end{align}

Having diagonalized $\hat R^2$, one can express the action of 
the $\hat R\inv$ operator on any function of definite angular momentum
$\Psi(z) = \sum_{n=0}^\infty f_n R_{n\ell}(z).$
In particular, the Schr\"odinger equation for the Coulomb problem,
$\left(\hat{P}^2/2m - k/\hat{R} \right) \Psi(p) = E\,\Psi(p),$
can be rewritten in the variable $z$ as
\begin{equation}
  \sum_{n=0}^\infty f_n
  \left[ \left(\frac{1+z}{1-z}\right)
    + \epsilon - \frac{2\xi}{\rho_{n\ell}}
  \right] R_{n\ell}(z) = 0 ,
\end{equation}
$\xi =\Delta x_{\min}/a_0$ being the ratio of the minimal length
$\Delta x_\mathrm{min}=\hbar\sqrt{\beta+\beta'}$
to the Bohr radius $a_0=1/km$, and $\epsilon=\xi^2 (E/E_0)$ 
the energy in units of the usual ground-state energy
$E_0 = -1/2a_0^2m$ times $\xi^2$.

Using the recursion relation for Jacobi polynomials as well as the 
orthogonality of the distance eigenfunctions $R_{n\ell}$, 
the Schr\"odinger equation is equivalent
to a recursion relation for the expansion coefficients,
\begin{equation}
f_{n+1} s_{n+1}\hat{a}_{n} + f_n t_n + f_{n-1} s_{n-1}\hat{a}_{n-1} = 0,
\label{recur}
\end{equation}
with $f_{-1}=0$, $s_n = 1-\epsilon + 2\xi/\rho_{n\ell}$,
\begin{align}
  t_n &= (2-s_n) - s_n  \frac{a^2-b^2}{(2n+a+b)(2n+a+b+2)}, \notag\\
\intertext{and}
  \hat{a}_n & =  -\frac{2}{(2n+a+b+2)}\notag\\
  &\quad\times\sqrt{ \frac{ (n+1)(n+a+1)(n+b+1)(n+a+b+1) }
    { (2n+a+b+1)(2n+a+b+3) }
  }.\notag
\end{align}

For a normalizable solution we must have
$
  \langle \Psi | \Psi \rangle = \sum_{n=0}^\infty f_n^2(\epsilon)
$ 
finite, thus $f_n$ should converge to zero.
A closed-form expression for this sequence cannot be determined. 
One can observe 
though that for large $n$ it asymptotically approaches 
\begin{equation}
  f_n \sim C_+\lambda_+^n + C_-\lambda_-^n 
  \qquad\text{with }
  \lambda_\pm=
  \frac{1\pm\sqrt{\epsilon}}{1\mp\sqrt{\epsilon}},
\end{equation}
$C_\pm$ being constants that depend on the minimal length through
$\xi$ and the energy eigenvalues through $\epsilon$. This allows one
to determine numerically the Coulomb spectrum, by imposing $C_+=0$.

As an independent check, we used two different algorithms. First, for fixed 
minimal length $\xi$, we imposed
$f_{n+1}/f_n=\lambda_-$ for sufficiently large $n$ and scanned for the values 
of $\epsilon$ for which the recursion \eqref{recur} 
gives $f_{-1} = 0$.
The contents following \eqref{fromHere}, 
and concluding with the first algorithm just 
described, represent results in unpublished work of Joseph Slawny.  We 
thank him for making these available to us prior to publication.

The second algorithm is more direct: for a given
minimal length $\xi$, we determined the values $\epsilon$ for which
$f_n$ converges to 0. The subtlety is that $C_+$ will never be 
represented internally as exactly zero, and the
term corresponding to it will eventually dominate our sequence. 
One can still identify 
the energy eigenvalues from the sign switch which occurs in $C_+$ 
and correspondingly in the large-$n$ behavior of $f_n$.

\begin{figure}
\includegraphics[scale=1.8]
  {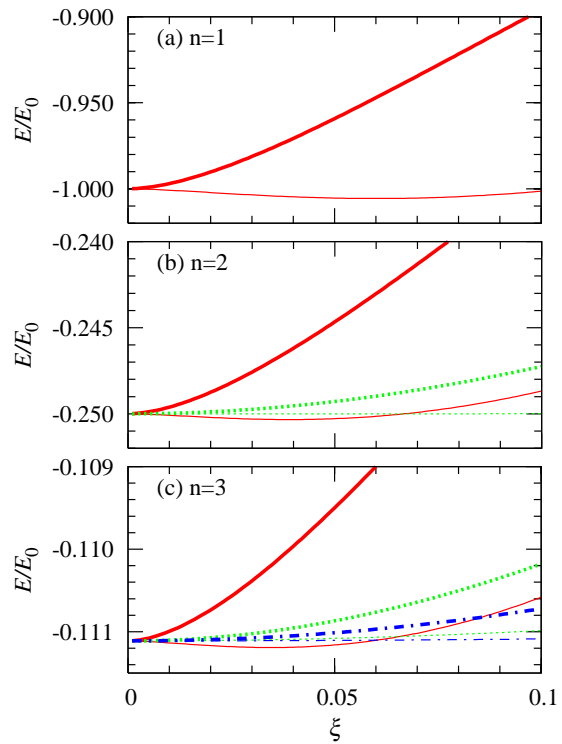}
  \caption{Plot of energy eigenvalues of the Coulomb potential 
    in units of the regular ground-state energy as a function of the minimal 
    length in units of the Bohr radius $\xi = \hbar\sqrt{\beta+\beta'}/a_0$.
    For principal quantum numbers  $n=1,2,3$,
    two extreme cases are represented: $\beta'=0$ (thick lines) and 
    $\beta=0$ (thin lines). Continuous line, $\ell=0$; dotted line, $\ell=1$; 
    dash-dotted line, $\ell=2$.
  \label{plots}
  }
\end{figure}

The algorithms yield consistent results, sampled in Fig.~\ref{plots}.
We can see that the degeneracy among different angular momentum
states is lifted: higher-$\ell$ states get smaller corrections. The
only exceptions are the $S$ states for $\beta'>2\beta$ and $D=3$. These
states start out with the lowest, negative correction for small
$\xi$, but cross the higher-angular-momentum levels as $\xi$ increases.
Another important remark, expected but not readily transparent in the
representation used, is that the energy values converge to the usual
result in the limit when the minimal length is taken to zero.

In order to get better insight into the observed behavior one can take
another approach. This is also needed because for the very small values of
the minimal length that interest us, i.e., several orders of
magnitude below the Bohr radius, the numerical convergence becomes
very slow and prone to rounding errors.

As mentioned before, the choice of momentum representation, 
while convenient, is not necessary. In relation \eqref{repr} 
one can use the ``pseudoposition'' representation
\begin{equation}
  \label{FakePos}
  \hat x_i = x_i, \qquad \hat p_i = \frac\hbar i\del{}{x_i},
\end{equation}
in which the position operators $\hat X_i$ are not diagonal, except for the 
limit $\beta=\beta'=0$. In this representation one can treat the $\beta$- and
$\beta'$-dependent terms as small, and use perturbation theory to deduce the
corrections to the energy spectrum. 

The operator $\hat R^2$ can be written as
\begin{equation} \label{R2-def}
  \hat R^2 = \hat r^2 + \xi^2 (\hat R^2)_{\xi^2} + \xi^4 (\hat R^2)_{\xi^4} ,
\end{equation}
where the first-order correction acts on the radial part of the
wave-function through
\begin{align}
\label{R2-1}
  \frac{(\hat R^2)_{\xi^2}}{a_0^2}  
  &= - 2\rho^2\del{2}{\rho} +[\eta(D-1)-3(D+1)]\rho\del{}{\rho} 
  \notag\\&\quad
  + \eta (2L^2+ D^2-D) -D(D+1), 
\end{align}
with $\rho = r/a_0$ and $L^2= \ell(\ell+D-1)$.

In the expansion of the inverse distance 
$ 
  \hat R\inv = \hat r\inv + \xi^2 (\hat R\inv)_{\xi^2} + O(\xi^4),
$ 
we expect on dimensional grounds that the first-order correction is a linear combination of
terms of form $(1/\rho)\partial_{\rho\rho}, (1/\rho^2)\partial_\rho$,
and $1/\rho^3$.  Indeed, substituting this form into $\hat R\inv \hat
R^2 \hat R\inv =\hat 1$ determines uniquely
\begin{align}
  \label{Omega}
  \frac{(\hat R\inv)_{\xi^2}}{a_0\inv} &= 
  \frac{[(2D-5)-\eta(2D-3)](D-1)-4\eta L^2}{4\rho^3}
\notag \\ &\quad
  +\frac{(3-\eta)(D-1)}{2\rho^2}\del{}{\rho}
  + \frac1\rho\del{2}{\rho}.
\end{align}
Thus, expressing the expectation value $\<(1/\rho)\partial_{\rho\rho}\>$ through the Schr\"odinger 
equation and $\<(1/\rho^2)\partial_\rho\>$ in terms of $\<1/\rho^3\>$,
the first-order corrections to the energy eigenvalues can be written as
\begin{multline}
  \frac{\xi^2}{a_0^2 m}
  \left\{
    \left[\frac{(D-1)(3\eta-1)}4 - \bar\ell(\bar\ell+1)(1-\eta)\right] 
    \!\Bigl<\frac1{\rho^3}\Bigr>
  \right.\\
  \left.
    + 2 \Bigl<\frac1{\rho^2}\Bigr>
    - \frac1{\bar n^2}\Bigl<\frac1{\rho}\Bigr>
    - \left.\frac{(D-1)(1-\eta)\rho^{D-3}\!}{2}
      [\Pi_{n\ell}(\rho)]^2\right|_0^\infty 
  \right\}\!,\\
\label{ExpVal}
\end{multline}
where   $\bar n = n + \frac{D-3}2$, 
        $\bar\ell = \ell + \frac{D-3}2$,  
and $\Pi_{n\ell}(\rho)$ is the unperturbed Coulomb radial wave-function.

A note of caution is needed here. While the expansion is apparently in
$\xi^2$, a quick calculation of higher-order terms confirms what is
expected on dimensional grounds, namely, that the expansion parameter
is $\beta/r^2 = \xi^2/\rho^2$.  
The $\xi$-quartic part in the expansion
\eqref{R2-def} of $\hat R^2$ contains terms of the type $\xi^4/\rho^2$
and $(\xi^4/\rho)\partial_\rho$.  
Therefore, the approximation
\eqref{Omega} for $\hat R\inv$ is no longer good for $\rho \lesssim
\xi$.  
In particular, in the actual operator $\hat R\inv$
there is no singularity at the origin.%
\footnote{This is quite
  general. Any $\hat P$-dependent commutation relation is expected
  to expand like \eqref{MLCR1D} and to exhibit this behavior.}

Let us estimate the error. 
The largest discrepancy between \eqref{ExpVal} 
and the actual value comes from the 
expectation value of $1/\rho^3$ calculated over the interval 
$[0,\rho_c\xi]$ on which the approximation \eqref{Omega} breaks down,
where $\rho_c \equiv r_c/a_0\sim 1$. 
For an angular momentum state $\ell$, this is of the order
\begin{multline}\label{error}
  \xi^2\int_0^{r_c\xi}\frac{R_{n\ell}^2(r)}{r^3} r^{D-1}dr 
  \sim \xi^2\int _0^{r_c\xi} r^{2\ell+D-4} dr \\
  \sim \xi^{2\ell+D-1}, \qquad\text{for $\ell>0$ or $D>3$.}
\end{multline}

For $D>3$ or $\ell \ne 0$, this contributes only a higher-order term; thus
it is safe to use \eqref{ExpVal}, and we finally arrive at
\begin{equation}
  \frac{\Delta E_{n\ell}}{E_0} = \frac{2 \xi^2} {\bar n^3}
  \left[\frac{(D-1)\left(3\eta-1\right)}
    {4\bar\ell(\bar\ell+1)(\bar\ell+\frac12)}
    +\frac{\eta+1}{\bar\ell+\frac12}-\frac1{\bar n}\right].
  \label{DeltaE}
\end{equation}

This expression generalizes the result of Ref.~\cite{Brau} for arbitrary
$\eta$ and $D$. For the particular case $D=3$ and $\eta = 1/3$ 
(i.e.,\ $\beta'=2\beta$) it reduces to the one obtained there. Moreover,
it is in excellent agreement with our numerical results.
\begin{figure}
  \includegraphics[scale=1.8]
  {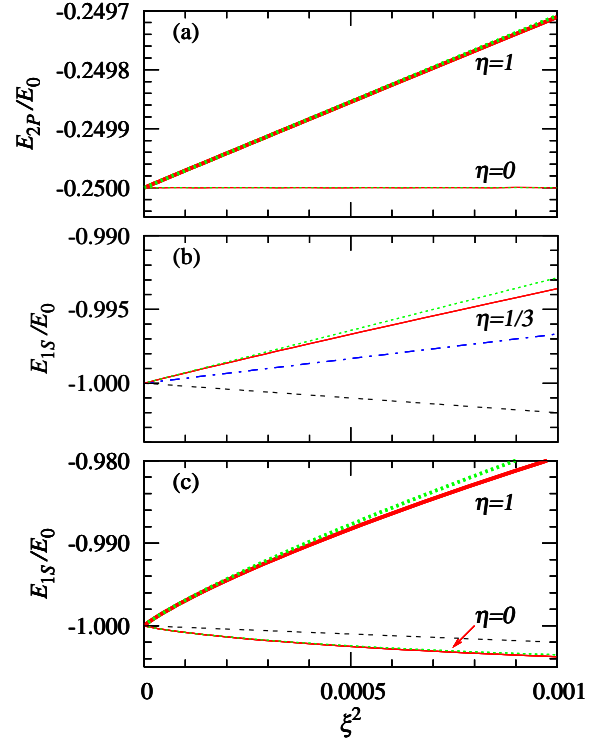}
  \caption{%
    Comparison of different results for 
      (a) 2$P$ states with $\eta=0,1$ and 
      (b), (c) 1$S$ states with $\eta = 0, 1/3, 1$. 
    Solid lines, numeric result;
      dotted lines, perturbative result; 
      dash-dotted lines, Ref.~\cite{Brau}; 
      double-dashed lines, Ref.~\cite{AY}. 
    The perturbative expression is given 
      for (a) by \eqref{DeltaE} and
      for (b), (c) by formula \eqref{DeltaEs}, with the coefficient 
      $C$ chosen such that it agrees with the numerics at $\xi^2=10^{-6}$.
  \label{compare}%
  }
\end{figure}
[See Fig.~\ref{compare}(a) for the case of the 2$P$ level.]

When $\ell=0$ and  $D=3$, the integral \eqref{error} is infinite. 
We can use only the part of \eqref{Omega} that can be trusted, 
i.e.,\ we have to cut off the expectation value integral $\<1/\rho^3\>$ 
at $\rho=\rho_c\xi$. The leading-order terms are
\begin{equation}
  \label{DeltaEs}
  \frac{\Delta E_{n\ell}}{E_0} 
  = \frac{4(3\eta-1)}{n^3} \xi^2E_1(\rho_c\xi) + C\xi^2 + O(\xi^3),
\end{equation}
where $E_1(\rho_c\xi)= -\ln \xi - (\gamma + \ln \rho_c) + O(\xi)$ 
is the exponential integral function. 
The coefficient of the $\xi^2$ term gets contributions 
from several parts. First, there are the remaining terms in \eqref{ExpVal}. 
Second, the actual value of $\hat R\inv$ is bounded on the 
interval $[0, \rho_c\xi]$, so by cutting off the integral at $\rho_c\xi$
we are neglecting another term of order $\xi^2$. 
Lastly, the exact choice of the cutoff value $\rho_c$ contributes another 
$\xi^2$ term. Because we do not know the exact form of $\hat R\inv$,
we cannot calculate analytically the second of these contributions.
When needed, $C$ can be determined numerically, by fitting relation
\eqref{DeltaEs} to the numerical results at a sufficiently low value
of $\xi^2$.%
\footnote{For $D>3$, $\< 1/\rho^3 \>$
    integrals are convergent, and the approximate solutions then
    agree nicely with the corresponding numerical results, even for
    $\ell = 0$.}%

When compared to the numerical results, the behavior of the energy as a
function of minimal length is nicely reproduced in
Figs.~\ref{compare}(b) and \ref{compare}(c).  
Our results disagree with Ref.~\cite{Brau}.
The difference is well explained by the neglect there of all but linear
terms in $\beta$. 
These terms critically affect the small-$r$ behavior of $\hat R\inv$, and
cannot be neglected.  Reference \cite{AY} arrives at a different
expression, which is independent of $\eta$. However, we could neither
account for the discrepancy nor reproduce those results.


We can finally set out to determine the constraint on the minimal length
$\Delta x_{\min}$ from precision hydrogen spectroscopy. A naive estimate,
obtained by imposing that the corrections are
smaller than the experimental error on the value of the hydrogen 1$S$-2$S$ 
splitting, gives $\Delta x_{\min} \gtrsim 300$~GeV 
(cf. \cite{Brau, AY}). Unfortunately, this estimate would be correct 
only if the measured value of the physical observable agreed with the 
theoretical prediction and the main source of error were the experimental one.
This is certainly not the case for the 1$S$-2$S$ splitting in hydrogen:
known to 1.8 parts in $10^{14}$, it is one of the most
precisely measured quantities today and is considered a \emph{de
facto} standard \cite{1S2S}. The value of the Rydberg constant is
determined using this measurement as an input, and thus the
theoretical uncertainty is orders of magnitude above the 
experimental one.

A better estimate is obtained by including contributions of the
(hypothetical) minimal length in the Lamb shifts. The strongest
constraint is expected from the 1$S$ Lamb shift, being the one
determined most precisely and getting the largest correction.  The
measured 1$S$ state hydrogen Lamb shift of $ L_{1S}^\text{expt} =
8172.837(22)$~MHz \cite{1Sexp} is larger than today's best
theoretical prediction $L_{1S}^\text{theor} = 8172.731(40)$~MHz
\cite{1Stheo} by about 5$\sigma$ experimental uncertainty.

If we attribute the discrepancy entirely to the minimal length 
correction to the 1$S$ state, the bound as a function of $\eta$, 
obtained using the first two terms in \eqref{DeltaEs}, 
is shown in 
\begin{figure}
  \includegraphics[scale=1.8]
  {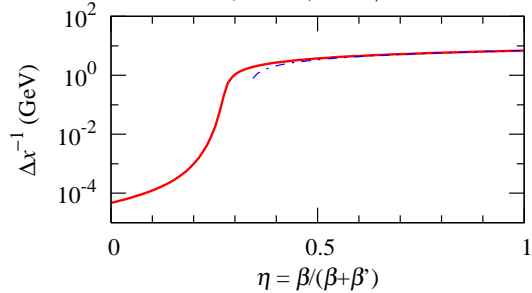}
  \caption{Constraint on the minimal length obtained as a function of $\eta$,
    including the two highest-order terms (continuous line) 
    and just the leading-order term (dash-dotted line).%
    \label{constraint}%
  }
\end{figure}
Fig.~\ref{constraint}.
It is 1.75~GeV for $\eta=1/3$ and increases to 6.87~GeV
for $\eta = 1$. Below $\eta = 1/3$, the constraint relaxes rapidly. 
Indeed, in this case the leading-order term in \eqref{DeltaEs}
is negative, and only the contribution from the next term can account 
for the observed difference. As a comparison, including only the 
leading term, we can obtain a bound only for $\eta>1/3$, with consistent 
results for $\eta\gtrsim 0.5$.

We should point out that the theoretical Lamb shift predictions 
are somewhat frail because of the uncertainties
in the proton charge radius \cite{2Ptheo}. 
These are the same order of magnitude as the ones discussed here;
thus one should consider the values in Fig.~\ref{constraint} 
rather as upper limits for the minimal length.
There is also the possibility of using  muonium spectroscopy, 
but the current limits are still weak for our  purposes. Details and
other implications for QED are under investigation.

\begin{acknowledgments}
The authors would like to thank F.~Brau, M.~Koike, J.~Slawny, and Y.~P. Yao 
for insightful discussions.
This research is supported in part by the US Department of Energy grant
DE--FG05--92ER40709.
\end{acknowledgments}


\end{document}